\documentstyle[psfig,conf_iap,10pt]{article}
\begin{document}
\heading{%
%
Clustering around QSOs at 1.0$<z<$1.6
%
} 
\par\medskip\noindent
\author{%
Sebasti\'an F. S\'anchez$^{1,2}$, J.I. Gonz\'alez-Serrano$^{1}$
}
\address{%
Instituto de F\'\i sica de Cantabria-CSIC, Avd. De los Castros S/N, 39005-Santander, Spain
}
\address{%
Dept. de F\'\i sica Moderna, Facultad de Ciencias, Univ. de Cantabria, Avd. De los Castros S/N, 39005-Santander
}
%

\begin{abstract}
Deep $B$ and $R$ band images ($R<$24 and $B<$25) have been obtained of
a field of $\sim$6$\times$6 arcmin around seven radio-loud quasars
with $1<z<1.6$. A statistically significant excess of {\it faint}
($22.5<B\le 23.5$ and $22<R\le 23.0$) galaxies was found, on scales of
$r<$160'' around the quasars
($>3\sigma_{N{\rm exp}}$). The galaxies have magnitudes
compatible with being at the redshift of the QSOs.

\end{abstract}
\section{Introduction}
Evidence has been accumulating over last decades that there is a link
between the quasar activity and overdensities of galaxies around the
host galaxy (\cite{yg87}, \cite{yg93} , \cite{HG98}, and
references therein).
Recent results, \cite{HG98}, have shown that there is an excess of
$faint$ galaxies around a sample of 31 radio-loud quasars at
high-$z$ ($z>1$). The magnitudes and colors of the excess galaxies are
consistent with a population of mostly early-type galaxies at the
quasar redshift. The size of this overdensity has not been completly
determined, beeing larger than $\theta\sim$100'' (due to the size of
the images, about 4'$\times$4').


We report on the results obtained from deep ($B<\sim$25 and
$R<\sim$24) CCD imaging survey of the fields around 7 B3VLA quasars,
in the redshift range $1.0<z<1.6$.  The images were obtained during
the night of 1997 March 13 at the 2.2m Calar Alto telescope, with a
final integration time of 1800 and 2400 seconds for the $R$ and $B$
band, respectively. The size of the images, $6'\times 6'$ have
allowed us to sample the density of galaxies to scales down to
$\sim$190''.


\section{Data processing and number counts}
We searched for objects in the $B$ and $R$ band images using the
SExtractor package. Our detection thereshold was of 5 connected pixels
above 2$\sigma$ level per pixel, which guarantees a detection up to
4$\sigma$. Simulations have been performed to determine the limits of
the validity of the method (detection+magnitude
determination+star-galaxy classification) showing that it is valid
down to $R\le$23.5-23.8 and $B\le$24.3-24.5. 

The $field$ number-counts and its variance have been obtained directly
from the images, measuring the number of galaxies in an anulus from
$r_{in}$=170'' to $r_{out}$=185'', and in a grid of 16 circular, not
overlapping, areas of $r<$35'' (centered more than 70''
from the quasars). The measurements were repeated for different ranges of
magnitudes, which define the number-counts distribution.

\section{Results}

Figure 1 shows the relative excess of galaxies averaged over the seven
fields, at both $B$ and $R$ bands. There is a significant excess of
galaxies around the quasars (5.52$\sigma$). The excess presents a
tendency with the magnitude range, being significant only for the
faintest magnitudes ($22\le R\le23$ and $22.5\le B\le23.5$), for any
angular scale (down to $\theta\sim$160'').

These results indicate most probably that the excess is not due to
intervening galaxies, since there is $no$ significant excess down to
$R<$21 and $B<22.5$ mag. The magnitudes ($R\sim$22.5 and
$B\sim$23.0), angular scale and number of galaxies of the excess are
compatible with beeing clusters or groups of galaxies at the redshift
of the quasars.




%
\begin{figure}
\centerline{\vbox{
\psfig{figure=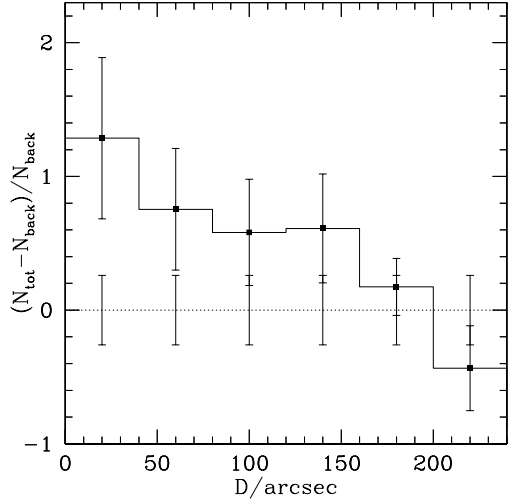,height=7.cm}
}}
\caption[]{Relative excess galaxies averaged over the seven quasar
fields and both bands.
}
\end{figure}




%
\vfill
\end{document}